\documentclass{iaus}
\usepackage{graphicx}

\title[large empirical libraries] 
{Analysis of stellar populations with large empirical libraries at high 
spectral resolution}

\author[Prugniel et al.]   
{Philippe Prugniel$^{1,2}$%
  \thanks{prugniel@obs.univ-lyon1.fr},
 Mina Koleva$^{1,3}$, Pierre Ocvirk$^4$, Damien Le~Borgne$^4$,
 Caroline Soubiran$^5$}

\affiliation{
$^1$Universit\'e Lyon~1,
Observatoire de Lyon, St. Genis
Laval, F-69230, France ; CNRS UMR5574;\\[\affilskip]
$^2$Observatoire de Paris, GEPI, F-75014, France;\\[\affilskip]
$^3$Department of Astronomy, St. Kl. Ohridski University of Sofia, BG-1164 Sofia, Bulgaria;\\[\affilskip]
$^4$CEA Saclay/Service d'Astrophysique, Gif-sur-Yvette Cedex, F-91191, France;\\[\affilskip]
$^5$Observatoire de Bordeaux, F-33270, Floirac, France;
}

\pubyear{2007}
\volume{241}  
\pagerange{1-2}
\date{30 Jan 2007}
\jname{Proceedings Stellar Populations as Building Blocks of Galaxies}
\editors{A. Vazdekis \& R. Peletier, eds.}
\begin{document}

\maketitle

\begin{abstract}
The stellar population models dramatically progressed with the
arrival of large and complete libraries, ELODIE, CFLIB (=Indo-US) and MILES
at a relatively high resolution.
We show that the quality of the fits is not anymore limited
by the size of the stellar libraries in a large 
range of ages (0.1 to 10 Gyrs) and metallicities (-2 to +0.4 dex).
The main limitations of the empirical stellar libraries are (i) 
the coverage of the parameters space (lack of hot stars of low metallicity),
(ii) the precision and homogeneity of the atmospheric parameters
and (iii) the non-resolution of individual element abundances (in particular
[$\alpha$/Fe]). Detailed abundance measurements in the large libraries, and
usage of theoretical libraries are probably the next steps, and we show that
a combination between an empirical (ELODIE) and a 
theoretical library (Coelho et al. 2005) immediately improves the 
modeling of ($\alpha$-enhanced) globular clusters.

\keywords{
galaxies: abundances; globular clusters: general;
}
\end{abstract}

\section {High precision analysis of stellar populations.}

The new empirical libraries that became available in the last 5 years 
allowed dramatic improvements in the analysis of stellar populations.
The ELODIE library (Prugniel \& Soubiran 2001, 2004; a new version, 3.1, will
be released soon) contains
2000 spectra of 1400 stars at high spectral resolution 
(0.55\AA $\approx$ R=10000  and R=42000) 
in the wavelength range 390 to 680 nm.
The CFLIB (or Indo-US) library (Valdes et al. 2004) has a lower resolution
(about 1\AA) but an extended wavelength coverage (346 to 946 nm). It
contains 1273 stars but suffers from a poor flux calibration. MILES
(S\'anchez-Bl\'azquez, 2006), finally,
has a slightly lower resolution (2.3 \AA) 
and narrower wavelength range (352 to 750 nm) but has been carefully 
flux-calibrated for 985 stars.
The coverage of the parameters space is comparable for the three libraries.

The increase of spectral 
resolution is a noticeable characteristic. In the past, population models
like Pegase (Fioc \& Rocca-Volmerange, 1997) were suited for multi-band 
photometry or low resolution spectra. 
With the new models, population analysis becomes sensitive 
to the depth and shape of the spectral features. As shown in Koleva et 
al. (2006) the optimal resolution, in term of observing time required to get
a given precision on the kinematics, age or metallicity, is of the
order of the physical broadening of the object. For giant elliptical galaxies
a R=400 is sufficient, but for a globular cluster R=40000 
would be an advantage. In practice, a resolution of R=10000 is well
suited for the usual setups available on the spectrographs (the model should
have higher resolution than the spectrograph in order to inject the 
instrumental broadening in the model prior to the analysis).
To benefit from the high resolution, methods performing full spectrum
fitting simultaneously of the kinematics and characteristics
of the populations have been developed (Ocvirk et al. 2006; Chilingarian
et al., this conference).

Koleva et al. (posters at this conference) have shown the remarkable 
consistency between models independently built with these libraries:
Pegase.HR using ELODIE and Vazdekis using MILES, and the reliability
of the determination of the characteristics of the stellar populations
using full spectrum fitting. 
Figure 1 gives an example of a fit to M67 
(observations from Schiavon et al. 2004) with Vazdekis/MILES and 
Pegase.HR/ELODIE. The quality of the fit is unprecedentedly good and
the parameters of the population match well the CMD determinations.
The residuals to the two fits (rms $\approx$ 0.5 \%) are very similar, 
and the same exercise repeated on other observations shows
that they are dominated by (i) the noise and errors in the observations
and (ii) physical features, like the misfit on Mg and CN
due to the different abundances in the clusters and in the library.

\begin{figure}
\centering
\resizebox{14.5cm}{!}{\includegraphics{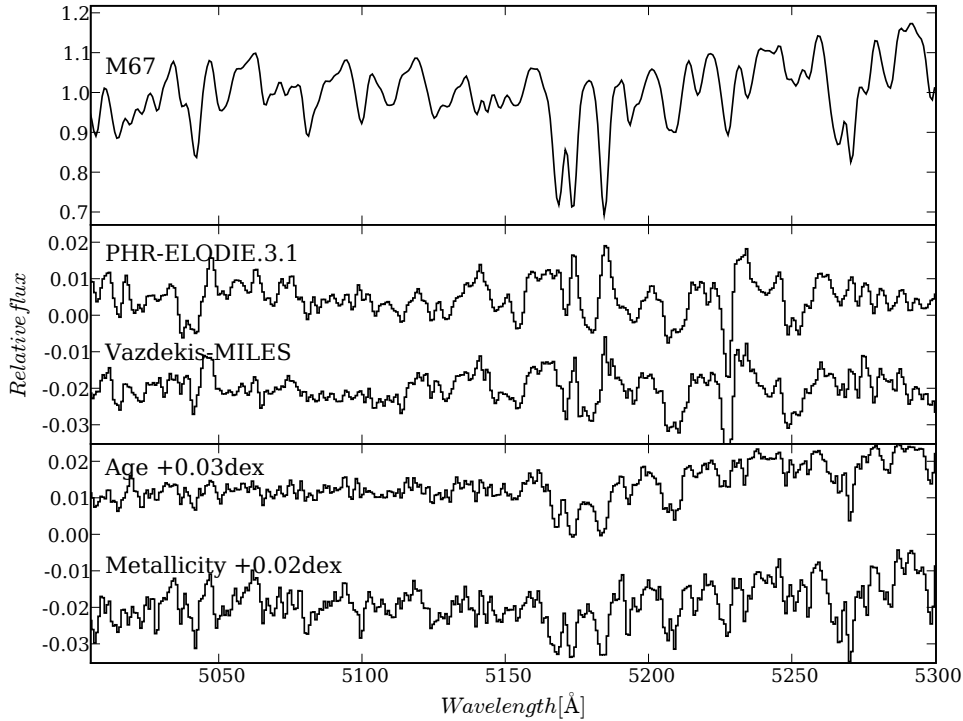}}
\caption {Example of population fit to the spectrum of M67. Top panel: M67
spectrum (from Schiavon et al. 2004) shown as reference. Middle panel:
residuals to the best fit with Pegase.HR and Vazdekis models, the vertical scale is magnified by a factor 10. Bottom panel: Effect of increasing the age by 0.03 dex (7.5 \%)
or the metallicity by 0.02 dex (5 \%) in the Pegase.HR-ELODIE.3.1 models.}
\label{Fig1}
\end{figure}

\section {Limitations of the present libraries}

\noindent {\it Atmospheric parameters of the stars.}
The effective temperature, surface gravity and metallicity of the stars in 
the libraries are essentially obtained from literature compilations.
Beside dissimilar quality and occasional errors, these
determinations result from various spectroscopic and photometric methods,
 based on
different stellar models, and despite the efforts to reduce the inhomogeneity, 
the global consistency, from the spectroscopic determination of the atmospheric
parameters to the computation of the isochrones in the plane of 
observed parameters, still deserves to be improved.

\begin{figure}
\centering
\resizebox{14.5cm}{!}{\includegraphics{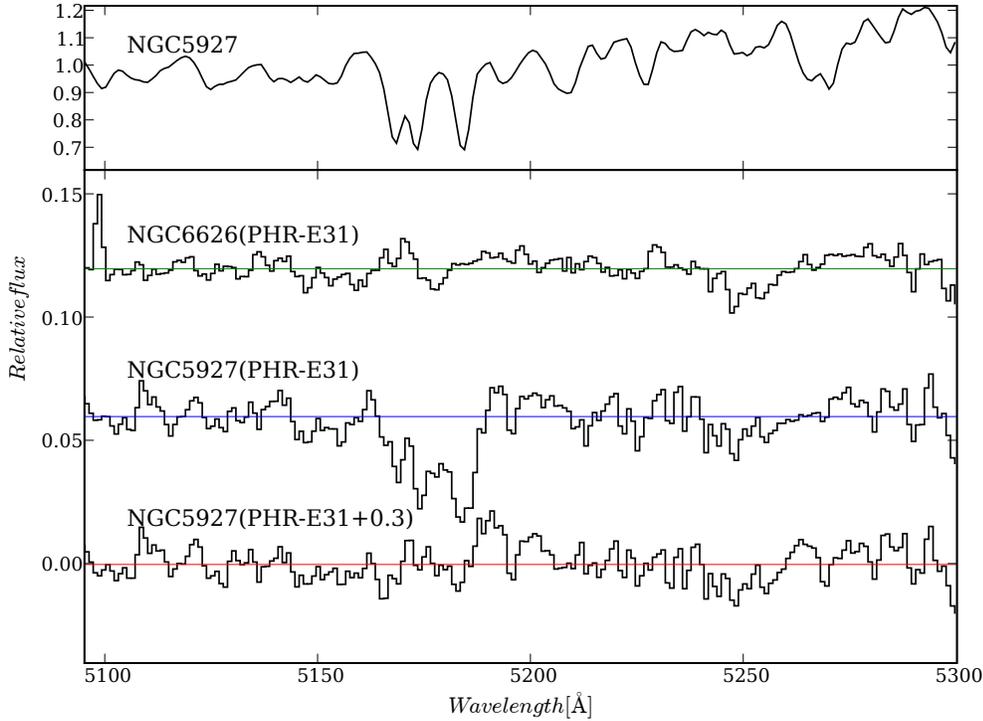}}
\caption {Population fit of the globular clusters NGC~5927 (high metallicity)
and NGC~6626 (low metallicity).
Top: Spectrum of NGC~5927 (from Schiavon et al. 2005) shown as reference. 
Bottom: (a) Residuals from the best-fit standard Pegase.HR-ELODIE.3.1 models 
for NGC~6626 and (b) for NGC~5927.
The latter shows important mismatch on Mg, due to Mg/Fe enhancement relative to 
the library.
(c) Residuals for NGC~5927 to the semi-empirical models where the ELODIE 
spectra are 
differentially corrected for [$\alpha$/Fe] abundance using Coelho et al. (2005).
The fit, made assuming [$\alpha$/Fe]=+0.3, is considerably improved.}
\label{Fig2}
\end{figure}

\noindent {\it Coverage of the parameters space.}
Lack of hot stars of low metallicity.

\noindent {\it Abundance of individual elements.}
The libraries have the characteristic abundance pattern of the solar
neighborhood, and therefore the models fail to reproduce the high metallicity
populations with over-abundant Mg, like elliptical galaxies
or bulge globular clusters (see Fig. 2). The classical solution to overturn
this limitation is to compute response functions of spectrophotometric
indices to individual elements (Korn et al. 2005).
However this approach does not exploit the high spectral resolution of 
the libraries.

\section {Next generation of models}

\noindent {\it ELODIE.4.}
We have recently assembled a new version of the ELODIE library (version 3.1) 
where the data reduction and determination of the 
atmospheric parameters have been improved. The spectral range (390-680 nm) 
has also been extended to the blue limit of the spectrograph.
To follow this version, we have selected from the ELODIE archive 
(Moultaka et al. 2004) 
all the stars present in the different libraries used for population 
synthesis and
we also included stars from sparse regions of the parameters space.
The new database consists of about 5000 spectra for over 2000 stars.
Using the ELODIE spectra at a resolution of R=42000 will help to produce an 
homogeneous
set of atmospheric parameters for a wide range of parameters, and to provide 
measurements of individual abundances and rotation.

\vskip 3pt
\noindent {\it Semi-empirical library with [$\alpha$/Fe] resolution.}
To overstep the limitation due to the unresolved [$\alpha$/Fe] abundance
of the libraries without loosing the advantage of the high spectral 
resolutions, the solutions may be (i) to build empirical libraries
with variable [$\alpha$/Fe], (ii) to use theoretical libraries or 
(iii) to adopt a mixed
approach where a theoretical library is used to correct an empirical one.
The latter possibility offers the advantage to cover the whole possible
range of [$\alpha$/Fe] at any metallicity, which would not be reachable
with a purely empirical library. This mixed approach avoids also the 
limitations of the theoretical libraries due to inaccuracy
or incompleteness of line lists, oscillator strengths and atmosphere models.
We present here a first attempt to generate such a semi-empirical library
where each wavelength point of the grid based on the ELODIE library is
differentially corrected to any  [$\alpha$/Fe] using Coelho et al. (2005).
It is assumed that the ELODIE library has the abundance pattern of the solar
neighborhood.

In Fig. 2, we show that while the fit to low metallicity globular 
clusters with the standard Pegase.HR models is acceptable because
the ELODIE library has a strong [$\alpha$/Fe] enhancement at low metallicity,
the mismatch of the Mg becomes important for high metallicity clusters.
Using the models built with the semi-empirical grid at an enhancement
[$\alpha$/Fe] $\approx$ +0.3, the residuals are canceled.

\vskip 3pt
\noindent {\bf Acknowledgements}
MK acknowledges the IAU financial support. PhP thanks FS.

\end{document}